# SOUND AND NOISE: PROPOSAL FOR AN INTERDISCIPLINARY LEARNING PATH

**Vera Montalbano,** *Department of Physical Sciences, Earth and Environment, University of Siena*

**Abstract**
A learning path is proposed starting from the characterization of a sound wave, showing how human beings emit articulate sounds in the language, introducing psychoacoustics, i. e. how the sound interacts with ears and it is transduced into an electrical signal for transmission to the brain. What is perceived as noise is presented and the concept is extended to physical measurements. The interdisciplinary teaching process is focused on active learning through activities at school and outside performed with an open source software allowing to record sounds and analyze spectral components.

## 1. Instructions

The sound can be studied in acoustic and correlated to other wave phenomena such as light. However, in order to define the noise, more information are required. In particular the concept of noise depends on human perception of sound, that is from psychoacoustics.

We propose a learning path that starts from the characterization of physics aspects of a sound wave and continues by giving some elements of phonetics, i.e. as human beings emit articulate sounds in the language. The next step is introducing psychoacoustics, that is how the sound interacts with our ears and in what way it is transduced into an electrical signal which is transmitted to the brain. Finally, what is perceived as noise is presented. The path ends with the extension of the concept of noise in physical measurements. The teaching process is focused on the relationship between related topics in different disciplines such as physics and biology. The learning process is favoured by activities in which students are encouraged to active learning. In particular, the characterization of sounds, phonemes and noise is proposed to students through activities to do both at school and outside by using an open source software allowing to record sounds with a computer and then to analyze the various spectral components. Many useful MM tools for sound and related topics are available and a selection of them can be used or proposed for further activities, but a particular care is given in designing measures in which students can choose which sounds are under examination. They are encouraged to compare different sounds from their environment, to analyze and evaluate them from the point of view of physics. Another aspect is the clarification of very important concepts in physics, commonly used in the practice of the laboratory, such as resonance and the signal-to-noise ratio.

The next section describes the motivation and purpose underlying this proposal and methodologies. In section 3, the sequence of activities in learning path is given outlining the relevance, interdisciplinary aspects and analogies. The role of Multimedia tools in favoring active learning is discussed. A pilot experience performed within the Italian National Plan for Science Degrees[1] is presented in section 4. Finally, some remarks and conclusions are given in last section.

## 2. Interdisciplinary path and active learning

The main purpose for proposing a learning path on sound and noise is that it links a classical topic in physics, sound, with an actual problem in our society, noise. The path correlates physics to student's day life and remedies their lack of awareness of the damage caused by noise. Moreover, it is a good example in which physics is useful for a better understanding of other sciences like biology, physiology, phonetics, electronics, etc. The best way to achieve this results is to design an interdisciplinary learning path.

According to H. H. Jacobs, an interdisciplinary path is obtained when a knowledge view and curriculum approach consciously applies methodology and language from more than one discipline to examine a central theme, issue, problem, topic, or experience (Jacobs 1989). Thus, it is possible to build a bridge between different disciplines and allow a better understanding of complex concepts utilized in science (in this case for example energy transport and transfer, resonance, transduction). Moreover, an interdisciplinary approach allows a synergy in acquiring a deeper knowledge of the world in which we live and, in this sense, it promotes an advanced scientific literacy.

---

[1] Piano nazionale per le Lauree Scientifiche, i.e. PLS
(see Montalbano 2012 for a survey on PLS actions and methodologies).

An effective way for pursuing these aims is active learning (Laws 1999, Niemi 2002, Meltzer 2012, Benedetti 2014). Students must be intellectually engaged and actively involved in their learning, and traditional instruction is usually failing to provide this engagement. On the contrary, a well-designed laboratory can achieve active learning through exploring and inquiry-based activities (Montalbano 2014a). In the laboratory context, multimedia tools can play a relevant role by enhancing the comprehension of some topics (e.g. transduction by means of applets as shown in fig. 3). Moreover, they can be used with mobile devices allowing students to perform measurements outside physics laboratory, at home or outdoors. Least but not less, many multimedia tools, like the one utilized for recording and analysing sounds in the following, are freeware so every teacher and every student can easily use it.

## 3. A learning path on Sound and Noise

The idea of an interdisciplinary learning path on sound and noise arose from a pilot experience in a deepening laboratory for motivated and talented students reported in the next section. Many activities are designed and tested with small groups of students. Disciplinary knots which were encountered, ideas for overcoming them and effective actions are merged in the proposal. In order to obtain an effective improvement in teaching/learning process a careful designing on all activities is suggested. A good way for promoting an effective design and a real interdisciplinary approach could be to organize an updating course for physics, mathematics, science teachers in-service in the school with the purpose of discussing and determining nature, degree of integration, scope and sequence of activities. The teacher's team should be empowered to shape and to edit the curriculum according to the students' needs.

The learning path starts from introducing waves and characterizing sound as a particular wave. Vary wave properties can be explored in lab, from energy transport and conservation to resonant systems. The next step is to understand how human being can produce sounds and which kinds of them are utilized in language. Moreover, noise is introduced analyzing day-life situations where students discover that a sound can be perceived like noise or not. Understanding how sound is perceived by the brain is necessary for arriving to a shared definition of noise. The last issues can be shaped directly on students interests (environment, pollution, health, music, electronic, advanced topic in science, etc.).

### *3.1. Waves and sound*

Waves are introduced in physics course but usually lab are centered on reflection, diffraction and interference of light. A way to study waves can be, starting from vibrating mechanical systems, to arrive to observe simple mechanical waves and to identify analogies and differences in both cases. A particular care is due because vibrating systems and waves are described by a very similar mathematics (harmonic functions) but few well-designed experiences in lab can clarify this delicate issue. Sound waves can be characterized and some common wave behaviors can be observed in a quantitative experiment, e. g. interference of sound waves, which allows to measure the speed of sound. An oscilloscope can be used for visualizing sound waves and allows to perform period and frequency measurements, verify the principle of superimposition, characterize harmonic waves and musical notes, compare them with known height (tempered scale), study sound beats and compare it with the analog phenomenon with light (Moiré fringes).

Energy considerations are relevant in defining waves, superposition effects, etc. In particular, resonance is a central topic for linking physics aspects to biological systems related to sound.

A very effective activity for understanding resonance (Montalbano 2014b) is the study of Chladni figures showed in figure 1. By using a speaker connected to a function generator, a resonant system can be obtained by placing a metal plate over it. When sound is resonant with the frequency of a normal mode of the plate, salt starts jumping, leaving from the vibrating surface and cumulating in fixed zones. The figures formed by salt (Chladni figures) depend on the shape of plates, materials, thickness and boundary conditions (existence of constrained points). In this case is very easy to recognize resonance.

Applets (see for example Physclips or Wikipedia) can be useful to figure out energy transport from sound wave to the plate and finally to the salt.

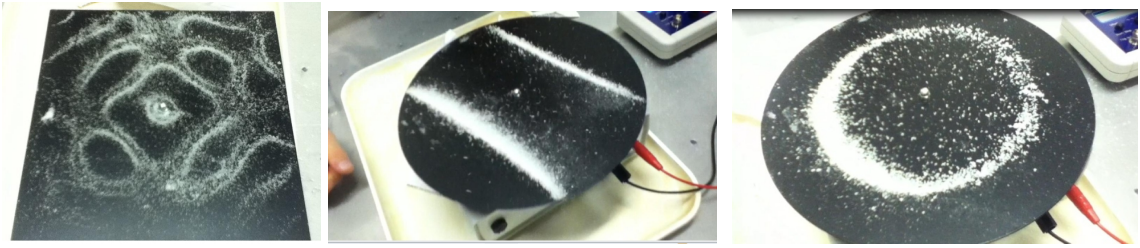

*Figure 1: A Chladni figure appears when a sound wave has the same frequency of a normal mode of the plate, thus the plate vibrates in a resonant way allowing a relevant exchange of energy from sound wave to the vibrating plate. Salt outlines the nodal positions of the normal mode.*

This part of the learning path can be realized by physics teachers, but a important role can be played by mathematics (harmonic functions, spectral decomposition, Fourier series, function in several variables) giving rise to an interesting interdisciplinary connection.

### 3.2. Sound production: phonetics
Sound is the basis of our communication system: The language. Phonetics studies the sound production: From the electrochemical signal processed by the brain to the sound produced by the human voice. The speech sounds are called phonemes. The position, shape, and movement of articulators or speech organs, such as the lips, tongue, and vocal folds allow to emit phonemes.
Speech organs (see fig. 2, picture in the middle) can be modellized as resonant boxes which amplifies only sounds with the resonant frequencies (as the one showed in fig. 2 on the left).
The first activity is to analyze some phonemes by using an open source software (e. g. Audacity, fig. 2 on the right). The software allows to record sounds by a personal computer or a mobile device. The recorded file contains a digital sampling of the analog electric signal in which sound energy is transduced by a microphone. Recorded files, as well as any other file that contains digitized sounds, can be read, modified, superimposed to other ones or decomposed into its spectral components by the software. Again, physics teacher is involved but collaboration with science teacher is unavoidable.

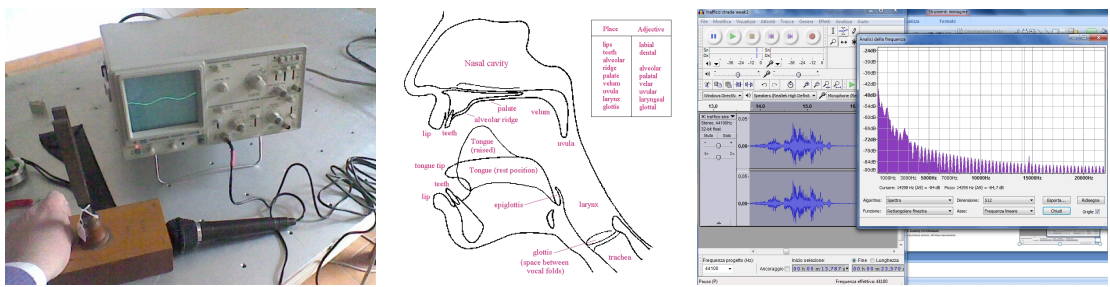

*Figure 2: On the left a resonant box for a note is showed together with the system microphone-oscilloscope used for obtaining a visualization of the electric analog signal in which the sound wave has been transduced. In the middle, a schematic picture of human speech organs is given (Fuaimeanna). On the right, some windows of the software are displayed.*

Students can characterize phonemes produced by themselves (simpler phonemes as a, e or ma, ba,da, mo, bo, lo,….). They can explore phonemes like an inquiry-based activity, posing questions (What happens if…..?) and checking hypothesis directly in the physical system (Let's try). They learn to use properly PC-software system and advanced topics can emerge and be discussed.

### 3.3. Noise and psychoacoustics
What is noise? In day life, we means mixed and unidentified sounds. For a well-defined concept of noise it is necessary to know how sounds are perceived by humans, i. e. how sound waves are detected by ears and transduced to an electrochemical signal ready the brain.
Energy aspects are relevant in psychoacoustic and they remain a key for understanding how sound is transduced by a vibrating mechanical system to a travelling wave in a viscous medium in the cochlea and finally in an electric signal by inner and outer hair cells (pictures from applets in fig.3). Moreover, the use of applets this time is not only useful but unavoidable.
The subject requires a deep interdisciplinary collaboration between physics and science teachers and heath aspects can be discussed (see fig. 3 on the bottom where hair cells damaged by noise can be compared to the normal ones).

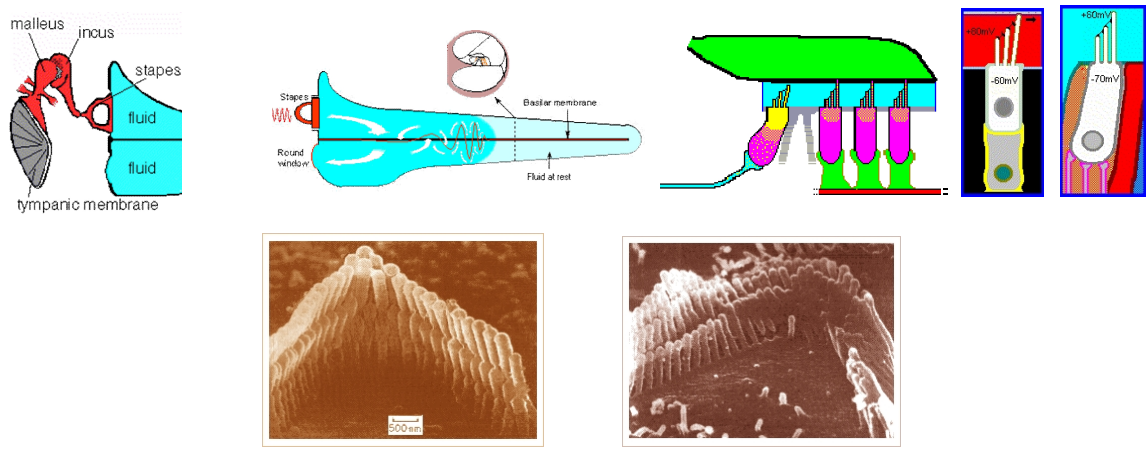

*Figure 3: On the top, vary applets useful for understanding signal transduction in the ear (from The Cochlea Site). On the bottom, normal and damaged hair cells (applets and electronic microscope images, from The Cochlea Site).*

### *3.4. Advanced issues on noise*
Students can deal with noise concept and related advanced issues, such as environment question (sound pollution, how it is defined, legal limit for noise, health aspects, how it is possible to reduce it, etc.). Students can comprehend that ear is an active transducer with well defined limit and their consequences on health.

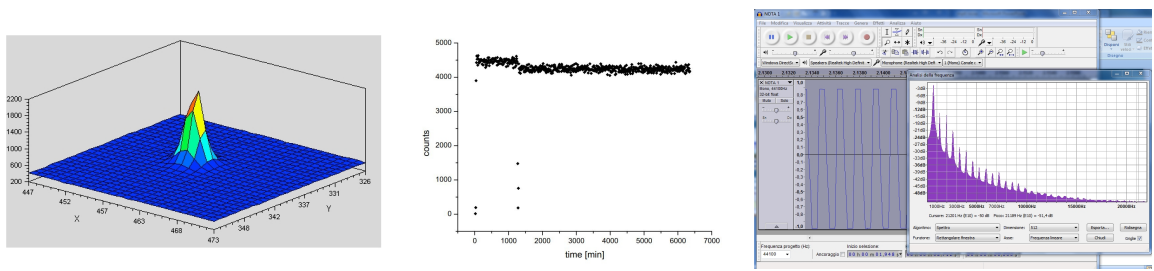

*Figure 4: Two measures with different S/N ratio (on the left, 4 in a measure on a fluorescente marker in a confocal microscope and, in the middle, 0.05 in a measure on weak radioactive source). On the right, a sound shows a cut off in the digital sampling.*

Finally, it is possible to turn back to physics and define signal-to-noise ratio (see fig. 4) and its importance in measurements. Students can compare S/N ratio for some physics measure and the case or ear for measured sounds. Moreover, they can explore sounds outside, in noisy environment, in the natural contest or other interesting sounds like white or colored noise and merge them with another sound for comparing the sensation from the ear. A limit in sampling can be found in the sound showed in fig. 4 on the left.

### 4. A pilot experience
Almost all issues proposed in previous section were designed and tested with students in a PLS laboratory. It was a deepening laboratory for motivated and talented students (3-4 students from scientific high school). The laboratory last for 3 years (4-5 h/month for a total 16-20 h/y) and students were very engaged and active. At the end of the PLS laboratory, a report was presented were many measures on different sounds were performed and analyzed by themselves using the open source software (see fig. 5).

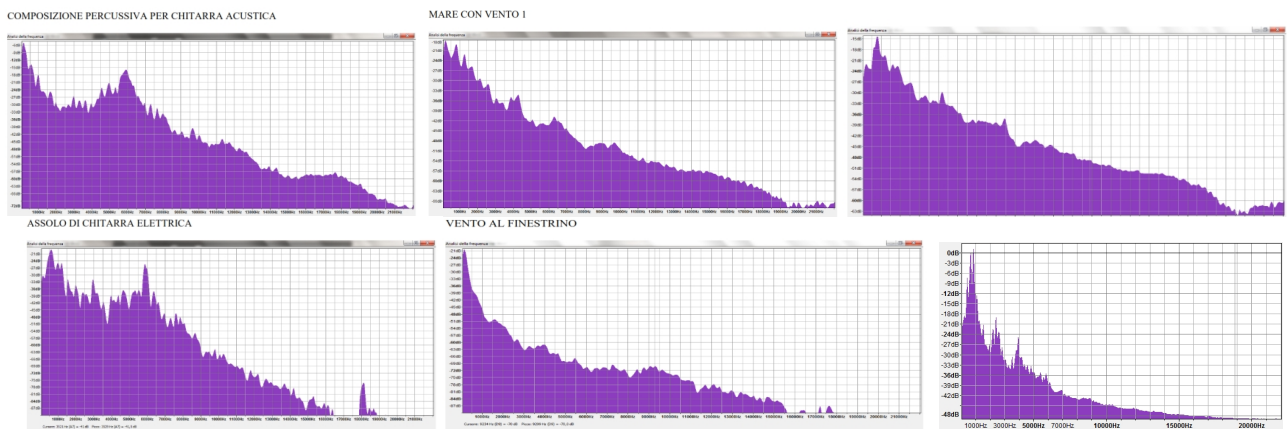

*Figure 5 Spectral analysis of sounds recorded and analyzed by students: acoustic guitar, wind on the sea beach, car stereo, electric guitar, wind on car window, phoneme A (from top on the left to bottom on the right).*

## 5. Remarks and conclusion

Pilot experience showed limits in an optional laboratory. Activities are too distributed in time. Analysis and final understanding are limited due to lack of discussion with teacher at school. In all activities in physics laboratory and in outdoor measurements students were very engaged and active but the optional aspect and the weak assessment which had no consequences at school render the lab not fully effective.

On this basis it has been proposed an interdisciplinary learning path at school with  physics, mathematics, science teachers involved. The activities can be adapted to different situations and enlarged following students' interests. The use of multimedia tools clarifies complex biophysics, implies an increasing of possible measures and makes students more independent in exploring the phenomena.